\documentclass[aps,prb,twocolumn,superscriptaddress,showpacs]{revtex4}
\usepackage{graphicx}

\begin{document}

\title{Comment on ``Extreme lowering of the Debye temperature
of Sn nanoclusters embedded in thermally grown SiO$_2$ by low-lying
vibrational surface modes''}

\author{Daniel B. Murray}
\affiliation{Mathematics, Statistics and Physics Unit,
University of British Columbia Okanagan, 3333 University Way,
Kelowna, British Columbia, Canada V1V 1V7}
\email{daniel.murray@ubc.ca}

\author{Avra S. Laarakker}
\affiliation{Mathematics, Statistics and Physics Unit,
University of British Columbia Okanagan, 3333 University Way,
Kelowna, British Columbia, Canada V1V 1V7}

\author{Lucien Saviot}
\affiliation{Laboratoire de Recherche sur la R\'eactivit\'e des Solides,
UMR 5613 CNRS - Universit\'e de Bourgogne\\
9 avenue A. Savary, BP 47870 - 21078 Dijon - France}
\email{lucien.saviot@u-bourgogne.fr}

\date{\today}

\begin{abstract}
This Comment corrects two separate errors which previously
appeared in the calculation of vibrational mode frequencies of
tin spheres embedded in a silica matrix.  The first
error has to do with the vibrational frequency of a free elastic
sphere.  The second error has to do with the effect on the
vibrational frequencies of a sphere due to its embedding in an
infinite elastic matrix.
\end{abstract}

\maketitle

In a recent article\cite{koops04}, Koops \textit{et al.} discuss the normal
modes of vibration of spherical tin nanoparticles embedded in a
glass matrix using an elastic sphere model.
However, the analysis errs in two respects in calculating the
frequencies of the vibrational modes.

The first issue has to do with the vibrational frequencies
of a free sphere, referred to here as the free sphere model (FSM).
Specifically, the frequency for
the breathing mode is often obtained incorrectly because of
the mis-application of a formula that is only applicable to
modes with angular momentum $\ell > 0$.  This error is
recurrent in the literature.\cite{kuok03,tanaka93}  Some authors
have since corrected them.\cite{erratumtanaka97,erratumkuok03}

\begin{figure}
  \includegraphics[width=\columnwidth]{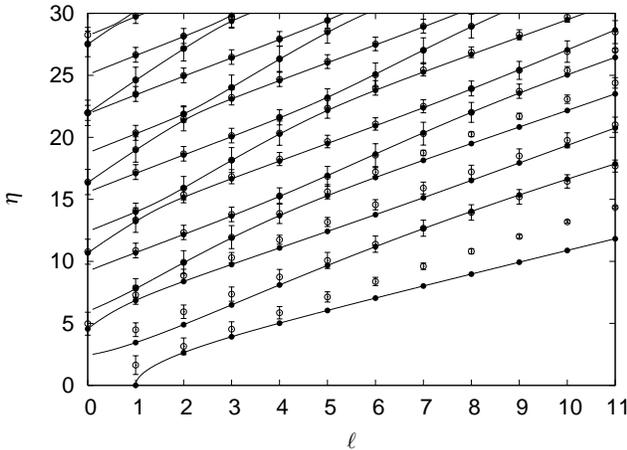}
  \caption{Corrected spheroidal FSM
  (solid line with filled circles) and CFM
  (open circles show real
  part, bars show half width at half maximum due to
  imaginary part of frequency)
  dimensionless
mode frequencies for a tin nanoparticle in a silica matrix
versus angular momentum $\ell$.
CFM matrix modes are not plotted
    }
  \label{fig1}
\end{figure}

\begin{figure}
  \includegraphics[width=\columnwidth]{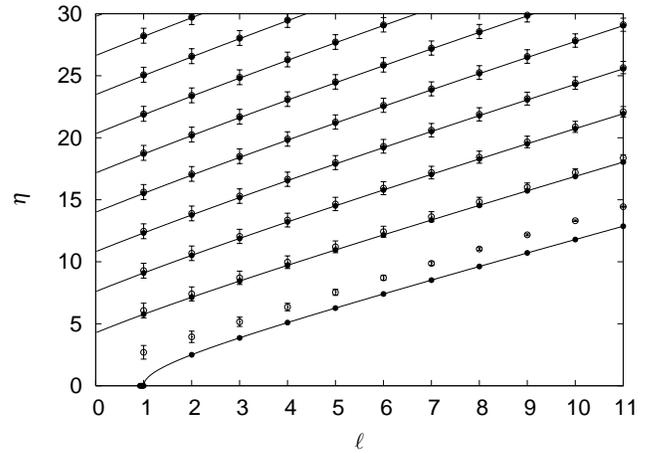}
  \caption{Corrected torsional FSM
  (solid line with filled circles)
   and CFM
  (open circles show real
  part, bars show half width at half maximum due to
  imaginary part of frequency)
     dimensionless
mode frequencies for a tin nanoparticle in a silica matrix
versus angular momentum $\ell$.
CFM matrix modes are not plotted
       }
  \label{fig2}
\end{figure}

The second issue has to do with the vibrational frequencies
of an elastic sphere embedded in an infinite elastic matrix,
referred to here as the complex frequency model (CFM) since
the transmission of vibrational energy out into the matrix
leads to temporally decreasing vibrational amplitude within the
nanoparticle, which is described with a frequency with a
nonzero imaginary part.
The original solution of the problem was correct\cite{dubrovskiy81}
and an explanation is available elsewhere.\cite{murrayPRB04}
This error results in incorrect estimates of the vibrational
frequencies of the acoustic phonon modes of the tin nanoparticle
in a glass matrix.

Figure~\ref{fig1} shows the corrected FSM and CFM
frequencies for the spheroidal modes of a tin sphere.
There are some additional CFM modes (not plotted in
the figures) with large imaginary parts.  These are called
matrix modes.\cite{murrayPRB04}
Koops \textit{et al.} and others have added straight
lines connecting the values of $\eta$ for consecutive integer
values of $\ell$, so as to attempt to indicate families of modes.
Instead, we have extended the FSM calculation of $\eta$ to
non-integer $\ell$.\cite{pr}  This could also have been
done for the CFM frequencies, but those lines are not shown.
The CFM frequencies apply to the case of a silica
matrix.  The speeds of sound for the tin nanoparticle
and the silica matrix are as in Koops \textit{et al.}
Figure~\ref{fig2} shows the corresponding frequencies
for the torsional modes.

Some past works\cite{ovsyuk96,tamura83b}
have calculated the vibrational frequencies of a sphere embedded
in an infinite matrix.  It is traditional to describe these
vibrations as ``confined phonons'', however vibrations of the
nanoparticle will be mechanically coupled to the glass matrix.
Consequently, energy in the form of outward travelling waves
will carry energy away from the nanoparticle, and result in
the amplitude of the vibrational energy decreasing with time,
implying exponential decay as well as both real and imaginary
solutions for the frequency.
Those earlier works found only real-valued frequencies for the
vibrations.  This does not make sense for modes which are
evidently damped.  The error in the earlier derivation is
due to unrealistic boundary conditions which were chosen
to be standing waves.\cite{ovsyuk96,tamura82} However,
as just mentioned, the physically appropriate boundary
conditions should be outward travelling waves
if the vibrational mode is considered to be confined
within the nanoparticle.

The same conclusion can be reached by using the Core
Shell Model (CSM) approach.\cite{murrayPRB04}  In CSM,
the nanoparticle of radius $R_p$ is surrounded by a
spherical matrix with finite outer radius $R_m$.  The
limiting case of $R_m \gg R_p$ is studied in order to
approach the situation of an infinite matrix.  All
normal modes of the system as a whole can be found,
and all frequencies are real valued.
In this situation, when the
macroscopic limit of a large matrix is taken, mode
frequencies are no longer discrete.  However, for
normalized modes, the squared displacement within the
nanoparticle interior peaks near frequencies which
correspond closely to the real parts of CFM frequencies.
In addition, the half widths at half maximum (HWHM) of these peaks
correspond closely to the imaginary parts of CFM frequencies.  As
a result, the understanding of the ``confined'' acoustic
phonons in the nanoparticle has been confirmed from two
quite different theoretical points of view.

In particular, it is important to note that the lowest
frequency vibrational mode of a nanoparticle is not the
breathing mode (the mode with purely radial oscillation).
Rather, it is a mode with angular momentum $\ell$ = 2.
With both of the errors now resolved, it is essential to
note that the lowest mode is the torsional $\ell$ = 2 mode,
whose displacement field has zero divergence, and that the
next lowest mode is the spheroidal $\ell$ = 2 mode.

D. B. M. acknowledges support from NSERC and the OUC
Grant-in-aid Fund.

\providecommand{\href}[2]{#2}\begingroup\raggedright\endgroup
\end{document}